\documentclass[a4paper]{article}
\usepackage{graphicx}
\usepackage{twocolpceurws}

\title{A Virtual Teaching Assistant for Personalized Learning}

\author{
Luca Benedetto \\ Politecnico di Milano, Milan, Italy \\ luca.benedetto@polimi.it
\and
Paolo Cremonesi \\ Politecnico di Milano, Milan, Italy \\ paolo.cremonesi@polimi.it
\and
Manuel Parenti \\ Politecnico di Milano, Milan, Italy \\ manuel.parenti@mail.polimi.it
}

\institution{}

\usepackage{booktabs} 

\begin{document}
\maketitle

\begin{abstract}
In this extended abstract, we propose an intelligent system that can be used as a Personalized Virtual Teaching Assistant (PVTA) to improve the students' learning experience both for online and on-site courses.
We show the architecture of such system, which is composed of an instance of IBM's Watson Assistant and a server, and present an initial implementation, consisting in a chatbot that can be questioned about the content and the organization of the \textit{RecSys} course, an introductory course on recommender systems.
\end{abstract}

\newcommand{\assistant}{Assistant}
\newcommand{\eg}{e.g.}
\newcommand{\ie}{i.e.}
\newcommand{\recsys}{\textit{RecSys}}

\section{Introduction}
Intelligent systems are extensively used in many domains, and they can bring some relevant advantages in education as well: indeed, they offer the opportunity to improve the learning experience and the quality of teaching, both in the case of online and on-site courses.
So far, diverse applications have been explored: for instance, some research focused on the usage of recommender systems for suggesting new learning content \cite{KM15} and to perform students' performance prediction \cite{Thai-Nghe10}.
Predicting students' performance is particularly important for e-learning, in order to improve retention and completion rates, which are one of the biggest limitations of online learning \cite{Hlosta17, Wolff13}.
Several works, such as \cite{duBoulay16, Goel16}, discussed the possibility of using virtual teaching assistants (VTA) in order to reduce professors' workloads: indeed, VTAs can make education much more scalable since students can solve most of their problems without asking the professors for help.
This extended abstract lies in this last branch of research: we propose a \textit{Personalized Virtual Teaching Assistant (PVTA)} for ``assisted learning'', which consists in helping students with a series of services such as personalization of content, recommendation of learning material and student engagement, as well as other services.
Also, we introduce an initial version of the PVTA, consisting in a chatbot - built leveraging the IBM's Watson Assistant - which is capable of answering students' questions about the content, the structure and the organization of the \recsys{} course, an introductory course on recommender systems.
The main differences from previous research consist in the personalization offered by the system and our focus on its architecture, showing how it is built, how it works and how it will be expanded.

\subsection{IBM's Watson Assistant}\label{watson}
The IBM's Watson Assistant\footnote{https://console.bluemix.net/docs/services/conversation} (called ``\assistant{}'' from now on) is offered by IBM as part of the Watson suite, an AI engine that provides several NLP services.
It is able to ``understand natural-language input and use machine learning to respond to customers in a way that simulates a conversation between humans'' and can be used to build virtual assistants.
IBM does not share with customers the details of the core NLP model, thus \assistant{} can be used only as a black-box; however, we can adapt it to any desired application domain by feeding it with additional data in order to enlarge the training set and fine-tune the model.
The training data we can feed the model with is made of objects belonging to two classes: \textit{intents} and \textit{entities}: \textit{intents} identify the goals that we expect a user to have while interacting with the system, while \textit{entities} affect the way in which \assistant{} reacts to a specific intent by giving it a context.
Once defined intents and entities, it is necessary to create the dialog flow in order to teach \assistant{} how it should answer different requests.

\section{Related Work}\label{rel_work}
In relation to this project, the most important work is the introduction of Jill Watson (JW) by A. Goel et al. \cite{Goel16}, a VTA somewhat similar to the PVTA proposed in this document.
However, although the authors showed the possible applications of JW, they never presented the details of the implementation and always considered it as a black-box; we aim at filling that gap, focusing on the architecture of our PVTA and explaining the role of the different components it is made of.
Also, there are some differences between the two systems: JW aimed at completely replacing human TAs, thus it had to deal with situations outside of the educational domain, which caused some problems that still have to be addressed \cite{Goel17}.
Our PVTA, instead, focuses on helping students in relation to the contents and the structure of the course, therefore we will not have to deal with that kind of issues.
Lastly, JW did not provide any kind of personalization, while we are building a personalized system.

\section{System Architecture}\label{sys_arch}
\begin{figure}[h]
\centering
\includegraphics[width=75mm]{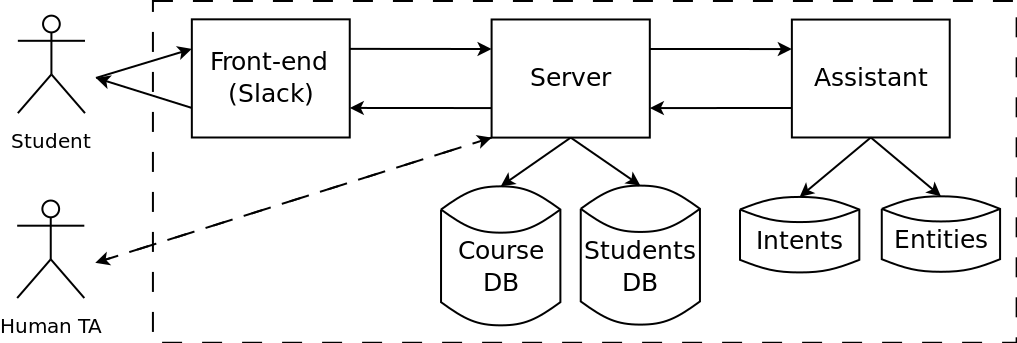}
\caption{Architecture of the PVTA. \label{fig:architecture}}
\end{figure}

As shown in Figure \ref{fig:architecture}, the PVTA is made of three main components: the front-end, a server and an instance of \assistant{}.
The server contains the data about the course and the students, while the instance of \assistant{} contains the intents and entities we defined.

\subsection{Front-end}
The front-end of the PVTA is a simple application that enables the user to interact with \assistant{} and does not perform any operations on the input data; for the first implementation we used Slack\footnote{https://slack.com/}.

\subsection{Server}
It is involved in different phases of the process: at the moment it performs up to three tasks for each question but this number will increase in the future, as we add new services to the PVTA.

\subsubsection{Preprocessing}
It is necessary if the server knows some information that \assistant{} is unaware of.
Watson has no long-term memory to store the context of the interaction with a student and we must leverage the server in order to do so.
As an example, if a student asks a question using the pronoun ``it'' referring to something he said in a previous question, the server has to modify the sentence in order for \assistant{} to understand such reference.

\subsubsection{Post-processing}
This is the analogous of preprocessing; for instance, when a student asks for the date of an exam, \assistant{} is able to understand what the user is interested in but it does not have access to the schedule (which is stored in the server). Thus, it sends an incomplete answer to the server, which fills the gap by adding the date and time of the exam.

\subsubsection{Students' modeling}
The server contains some intelligence as well: it collects data about students' behavior and clusters them in different groups using as similarity the intents and the entities they searched for.

\subsubsection{Interaction with a human TA}
Figure \ref{fig:architecture} shows that the server can interact with a human TA: \assistant{} assigns a confidence level to each answer and, if that is too low, the server forwards the question and the proposed answer to the human TA.
He checks the proposed answer, possibly corrects it and sends it back to the server, which forwards it to the student and inserts the correct question-response pair in the training set (stored in \assistant{}).

\subsection{\assistant{}}
It is responsible for the NLP-related tasks of the VTA: it receives the question (possibly preprocessed) and provides an answer (which might require post-processing).
In order to build a system capable of working in the educational domain, we had to define the \textit{entities} and the \textit{intents} related to such domain, as well as all the technical terms specific to the \recsys{} course.
Each entity does not represent a unique concept, but a group of concepts; also, each concept might be referred to with different synonyms.
So far, we defined 50 intents and 20 entities (more than 170 different concepts) but this number is likely to increase in the future, as we add new functionalities.

\section{Conclusion and Future Work}\label{future}
This document proposed an architecture for a PVTA capable of providing several services to students of online and on-site courses, moving towards the goal of ``assisted learning''.
Also, we introduced an initial implementation of such system, consisting in a chatbot capable of answering the questions of students enrolled in a course about recommender systems.
This chatbot is only a small part of the PVTA we propose, and we are working on new services to be implemented in the system.
We are working on ways to continuously enrich the set of \textit{intents}, \textit{entities} and rules by monitoring the interactions between students and the PVTA: by means of knowledge extraction algorithms we aim to keep fine-tuning and improving the model even after deployment.
We also plan to implement student engagement: looking at students' behavior, the PVTA can understand which are the students at risk of dropping-out and the ones not satisfied with the course; then, it can proactively intervene or send a warning to human TAs.
Another aspect we are focusing on is analyzing whether it is possible to reduce the usage of the IBM's Watson Assistant and perform a bigger part of the NLP-related tasks in the server.
Indeed, we have two objectives in mind: implementing some NLG (natural language generation) in the server in order to overcome the biggest limitation of \assistant{}, which is the impossibility to generate answers, and - as a second step - exploring the generation of personalized answers.
Lastly, further work is focusing on the possibility of personalizing learning material, recommending different contents depending on the student and its interactions with the PVTA.


\bibliographystyle{alpha} 
\bibliography{bibliography}

\end{document}